# Newtonian explanation of galaxy rotation curves based on distribution of baryonic matter

Konstantin Pavlovich, Alex Pavlovich & Alan Sipols

June 9, 2014

360 W Illinois St., 8F, Chicago IL 60654, USA
email: apavlovi@umich.edu

ABSTRACT

Circular velocities of stars and gas in galaxies generally do not decline in accordance with widely expected Keplerian fall-off in velocities further from the galactic nucleus. Two main groups of theories were proposed to explain the supposed discrepancy—first, the most commonly accepted, is the suggestion of the existence of large non-baryonic dark matter halo, and, second are theories advocating some modification to the law of gravity. So far however, there is no empirical evidence for either dark matter or modified gravity. Here we show that a broad range of galaxy rotation curves can be explained solely in accordance with Newton's law of gravity by modeling the distribution of baryonic matter in a galaxy. We demonstrate that the expectation of Keplerian fall-off is incorrect, and that a large number of likely galaxy mass distribution profiles should in fact produce flat or accelerating rotation curves similar to those observed in reality. We further support our theoretical findings with the model fit of 47 real galaxies' rotation curves, representing a broad range of galactic types and sizes, and achieving correlation of expected and observed velocities of over 0.995 for all cases. Our results make theories of exotic dark matter or modified gravity unnecessary for the explanation of galaxy rotation curves.

INTRODUCTION

Since Fritz Zwicky's observations[1] of galaxy rotation curves in 1930s astronomers were puzzled by the fact that circular velocities of stars and gas further from the nucleus of the galaxy generally do not decline in accordance with widely expected Keplerian fall-off, similar to the decline in the Solar system[2]. Multiple studies[3,4,5,6,7] confirmed that galaxy rotation curves are mostly flat, with some galaxies showing modestly declining[8] and some accelerating circular velocities further away from a nucleus. There are two main groups of theories proposed to explain these phenomena. First, the most commonly accepted, is the suggestion of existence of large non-baryonic dark matter galactic halo[9,10,11,12]. Second, there are theories advocating some modification to the law of gravity, with Modified Newtonian Dynamic[13,14,15] and Scalar–Tensor–Vector Gravity[16,17], being top contenders. In spite of intensive search, all attempts to detect non-baryonic dark matter failed. There is also no empirical support for modifications to the law of gravity.

Most of the research in the past approached the problem of galaxy rotation curves assuming spherically symmetric gravitational field due to a point source, using the gravity law in the form $F = G M m / R^2$ inside the galaxy disk with the entire galactic mass as a parameter. However, gravity law in this canonical form is only appropriate for two point masses or spherical objects, and only outside of the sphere's radius. We believe it is incorrect to apply it in this form to the entire galaxy mass for calculations within the galaxy disk, where gravitational forces have an integral effect from multiple directions. In our opinion, this oversimplification created expectation of Keplerian fall-off, which, in turn, led to the rise of theories of non-baryonic dark matter or modified gravity to explain the supposed discrepancy.

Kepler's laws are applicable for a system with a massive star in the center orbited by a few planets with immaterial mass. In such system, one can ignore mutual gravitational pull of planets, although even in this simple case Newton's adjustments to Kepler's 3rd law are required for precise calculation. Once we increase the number of objects in a system to billions, especially if the aggregate mass of these objects exceeds the central nucleus mass, Kepler's laws become inapplicable. Galaxy is this type of system—the mass of a central black hole is several orders of magnitude lower than the mass of a galaxy. Such system should be viewed as complex and uneven swarm of objects, which behavior can only be analyzed via integration of mutual gravitational forces between all those objects.

INTEGRATION OF GRAVITATIONAL FORCES IN A GALAXY

To calculate a cumulative gravitational force exerted upon any selected point in a galaxy, we have to integrate gravitational forces between selected point and all other points within a galaxy. In a two-dimensional[1] model of a galaxy the gravitational attraction force between random point A on x-axis with mass $M_A$ at a distance *r* from the center (Figure 1) and random point B(x,y) with $M_B$ is determined by the gravity law as:

$F_A = G\, M_B\, M_A / (AB)^2$   (1)

**Figure 1 | Theoretical illustration of gravity forces in a galaxy**

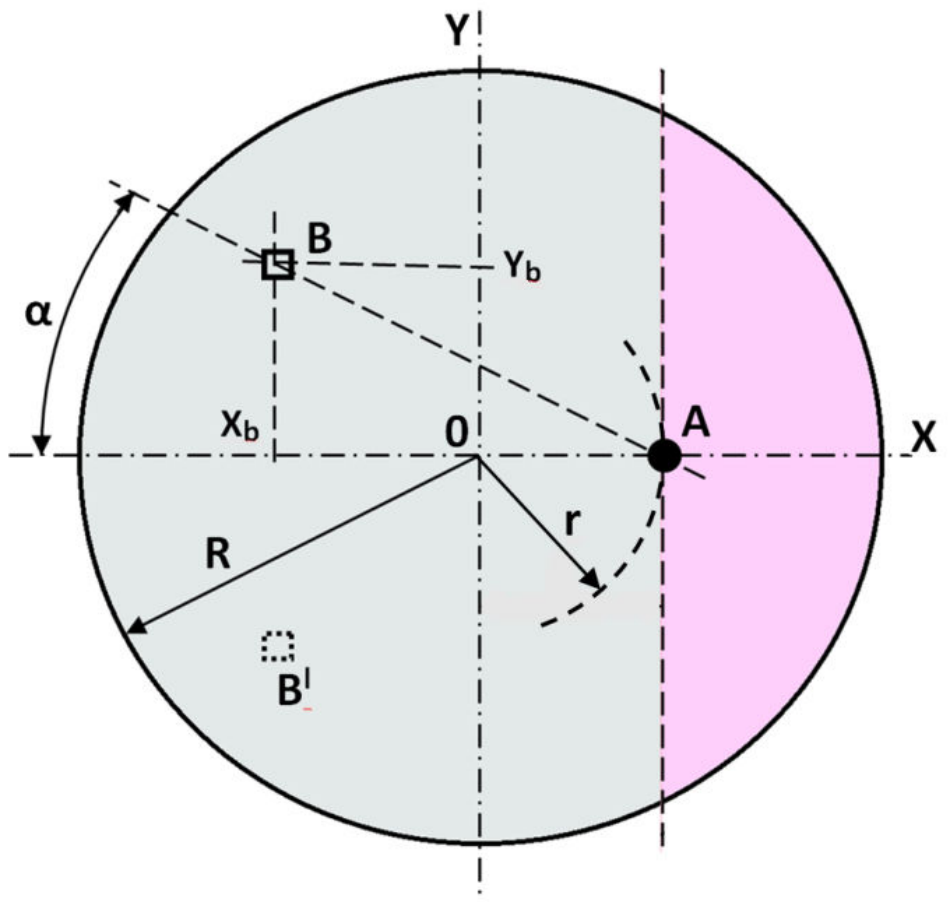

The vector of this force is at an angle α to the X-axis (Figure 1). The horizontal component $F_A\cos(\alpha)$ generates a centripetal gravity force and vertical component $F_A\sin(\alpha)$ generates a tangential gravity force. In general, the tangential component is significantly weaker than the centripetal as the gravitational pull of the mass above and below the X-axis largely cancel each other out. In the case of axisymmetric systems the tangential component $F_A\sin(\alpha)$ is completely compensated by an equal force $-F_A\sin(\alpha)$ from a mirror point B'. Integrated over galactic disk area D with radius R, the centripetal and tangential components of the gravity force are calculated as:

$$\text{Centripetal force } F_{Ac} = G\, M_A \iint_D \frac{M(x,y)\cos(\alpha)}{(AB)^2}\, dx\, dy \quad (2.1)$$

$$\text{Tangential force } F_{At} = G\, M_A \iint_D \frac{M(x,y)\sin(\alpha)}{(AB)^2}\, dx\, dy \quad (2.2)$$

Where $AB^2 = (r - x_b)^2 + y_b^2$;  $\text{Cos}(\alpha) = (r - x_b)/(AB)$; $\text{Sin}(\alpha) = y_b/(AB)$;

[1] The two-dimensional simplification of the model is warranted given that distances in the galactic plane are significantly greater than the distances perpendicular to the plane. Thicker profile in a bulge can be accounted for in a two-dimensional model by increased mass density in the center.

Given that centripetal force $F_{Ac}=M_A V_A^2/r$ we find orbital velocity $V_A$ as:

$$V_A = \sqrt{G\, r \iint_D \frac{M(x,y)\,(r - x_b)}{(AB)^3}\, dx\, dy} \quad (3)$$

Except for the uniform mass distribution case, integral (3) is unsolvable with direct methods for even simple functions of mass distribution M=f(x,y). However, by breaking the galaxy disk into a sufficiently large number of small areas as a matrix, we can approximate this integral with a sum:

$$V_A = \sqrt{G\, r \sum_{i=1}^{n} \frac{M(x,y)\,(r - x_b)}{(AB)^3}} \quad (4)$$

This approximation of integral (3) with a sum (4) is a crucial step. Some earlier researchers[18,19] acknowledged the integral (3), yet largely resorted to models with spherically symmetric gravitational field due to a point source, not reflecting galaxy disk structure and ignoring the impact of cumulative gravitational pull from low mass density galactic periphery. As we demonstrate below accounting for galaxy disk structure and peripheral mass outside of the measurement orbit makes a profound impact on galaxy rotation curves.

## GALAXY MASS DENSITY MODEL

In order to solve equation (4) for a large number of galaxies of various types and sizes, we created a galaxy Mass Density Model (MDM). The model is based on N-by-N matrix in which each cell represents a region of space scaled automatically to selected galaxy size. Setting matrix resolution at 41 by 41 enables approximation of the integral (3) with error tolerance of no more than 0.35%, which for practical purposes is sufficient to model key features and behavior of galaxy rotation curves. In principle, however, it is easy to make this model as granular as desired. Each cell in the matrix is ascribed a nominal mass density number from zero to 100, where zero is vacuum and 100 the densest area (Figure 2). The mass of each cell is then calculated as a total galaxy mass multiplied by respective cell nominal density number and divided by the sum of nominal density numbers of all cells. Calculating gravity forces between selected cell on the axis of measurement and all other cells we solve equation (4). By repeating this procedure for other cells along the axis of measurement we determine rotation curve for selected mass distribution and axis.

**Figure 2 | Galaxy mass density model setup**

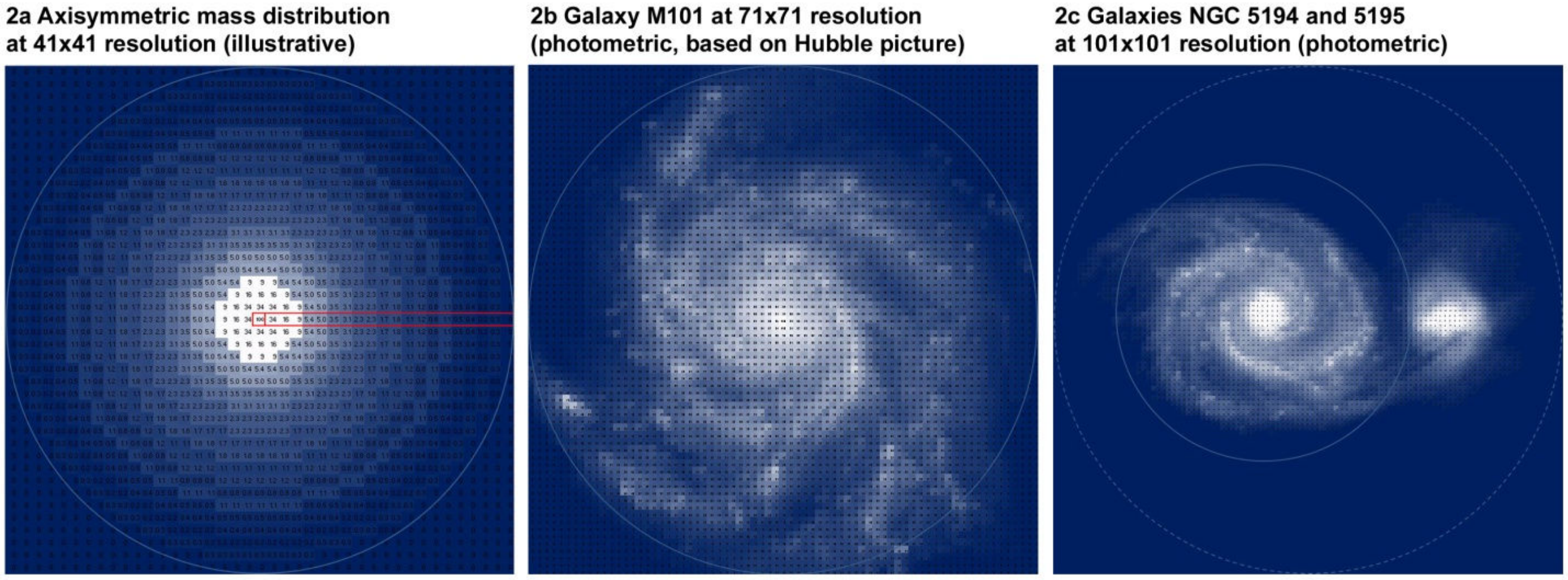

For an axisymmetric mass density distribution the whole matrix can be setup with a one dimensional array from nucleus to the edge of the galaxy (Figure 2a). Importantly, axisymmetric model setup allows us to run MDM in both directions—from mass distribution to rotation curve, and from rotation curve to implied axisymmetric mass distribution. As we demonstrate below, such simple axisymmetric setup is largely sufficient to explain general behavior and shapes of galaxy rotation curves even for spiral and other non-axisymmetric galaxies. This is due to the fact that in most cases centripetal forces are significantly stronger than tangential forces generated by non-axisymmetric structures such as spiral arms. When necessary however, the two dimensional functionality of our matrix enables us to model any non-axisymmetric distributions including all types of irregular features such as spirals and bars (Figure 2b) and account for tangential gravitational forces. It also allows us to utilize galaxy photographic data to infer approximate mass distribution and estimate circular velocities for various axes for galaxies with non-axisymmetric mass distributions.

To translate galaxy photograph into an approximate mass density distribution matrix, it is possible to measure average relative luminance for each rectangular area of the source image corresponding to a target matrix cell, and correct resulting luminance matrix values for sky background light and the mass density-to-relative luminance ratio (MD/RL). The MD/RL adjustment function can be generalized from correlations between the distribution profiles of relative luminance and mass density, evaluated on a sample of (a) galaxy images with known view angle and (b) mass density distribution estimates generated by MDM from rotation velocity inputs. For conversion of edge-on to face-on 3D projection the matrix can be further transformed taking into account galaxy spatial orientation parameters. Lastly, automatic heuristic correction algorithm and/or precise manual override can be applied to those individual cells where value is likely distorted by objects present on the image but known or suspected to be foreign to the galaxy.

## SHAPES OF GALAXY ROTATION CURVES IN RELATION TO MASS DISTRIBUTION

To explore theoretical behavior of galaxy rotation curves depending on a range of axisymmetric mass distributions we used a model of a hypothetical galaxy with a mass $10^{11}$ M☉ and a radius 20 kpc. We setup several axisymmetric mass density distribution profiles as a function of a distance from the nucleus $1/(x+1)^k$, where *x* is the distance (1, 2, ..., n kpc) and *k* is a real number (Figure 3b). As we can see (Figure 3a), the rotation curve flattens rapidly with even very low mass density at a galactic periphery. At k=1.5, which implies 1% periphery to center density ratio, the rotation curve is practically flat. For galaxies with relatively denser periphery at k<1.4 (implying edge to center density ratio above 1.4%) we should expect accelerating circular velocities further from the nucleus. Of course, galaxy mass distribution does not need to follow a specific function. Thus, for illustrative purposes we also constructed a hypothetical "Ideal Flat Curve" with the circular velocity set at a constant 178 km/sec at every point of the curve (Figure 3; red dotted line) and using MDM calculated implied mass distribution required to produce this rotation curve. This hypothetical Ideal Flat Curve implies even lower 0.2% periphery to center density ratio. It requires faster initial density drop from the nucleus, followed by slower density decline in the galactic disk. In general, Keplerian decline in velocities can only be achieved with almost instant density drop (Figure 3; green dashed line) which implies practically all galaxy mass in the center. Density distributions around the Ideal Flat Curve and up to k=1.5 produce approximately flat rotation curves. Density profiles above this area generate accelerating rotation curves.

**Figure 3**
**Expected rotation curves of a hypothetical galaxy depending on mass density profile**

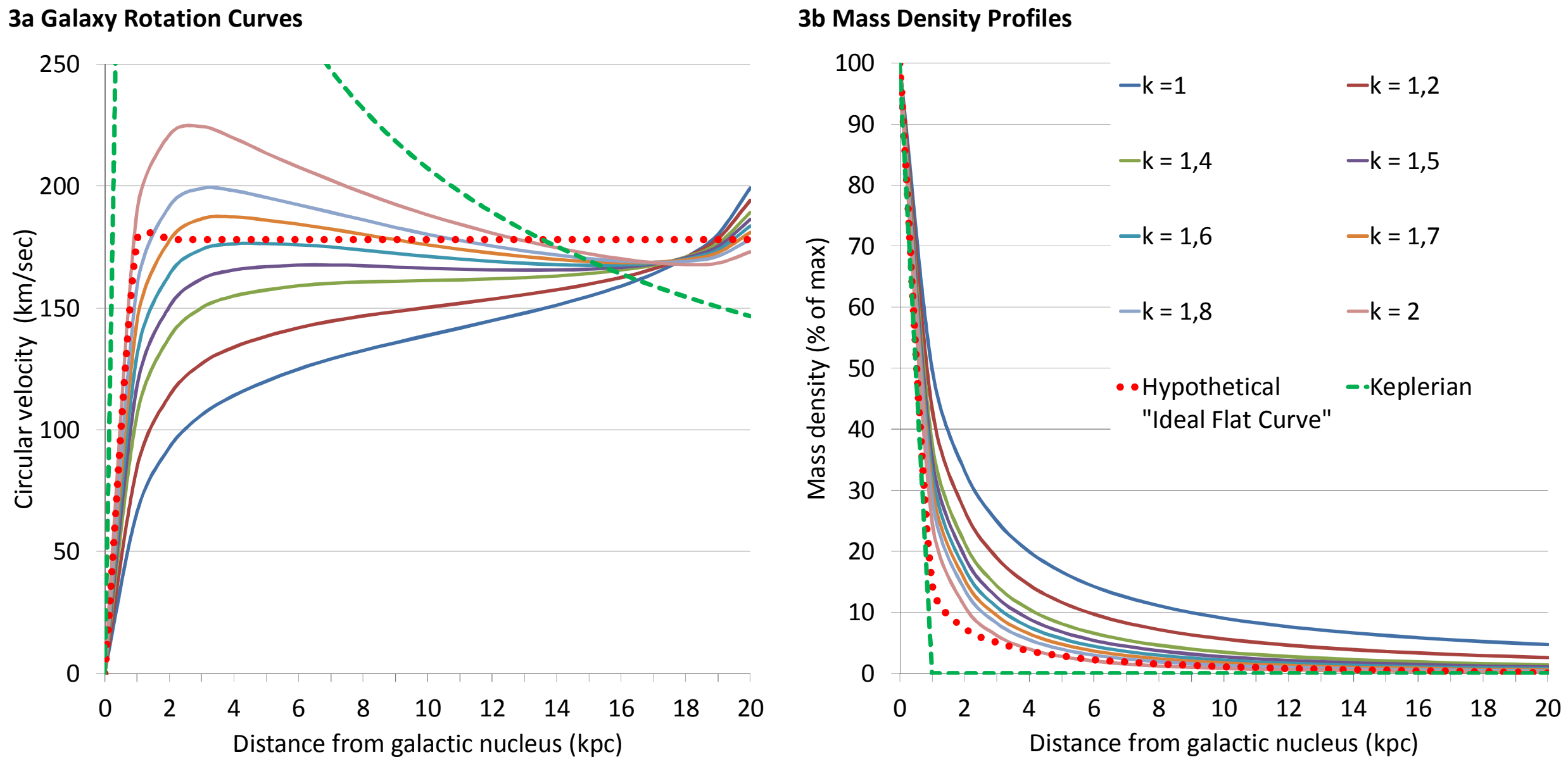

Conceptually, there are two reasons why very low peripheral density is sufficient to produce flat and even accelerating rotation curves. First, galactic periphery covers much larger area than the bulge and even at a low density cumulative mass of the disk become comparable with that of the bulge. Second, a low mass close to the point of measurement can exert more significant gravitational force locally than much heavier but distant galactic nucleus. The conclusion is inescapable—a system with mass distribution resembling that of any typical galaxy should in fact produce a rotation curve similar to those we observe in reality, without exotic dark matter, but strictly in accordance with known law of gravity.

## MODEL FIT TO ROTATION CURVES OF REAL GALAXIES

To test our Mass Density Model on real galaxies we selected a sample of 47 galaxies covering twelve galaxy types (Sa, SAb, SBc, etc.), broad spectrum of shapes of rotation curves and velocities, and a wide range of galactic diameters from 2kpc to 80kpc. Using MDM, we calculated galactic mass and its distribution solving for the best fit to respective galaxy rotation curve and diameter. For 46 of 47 galaxies MDM automatically achieved extremely close fit with correlation over 0.995 between observed velocities and those generated by the model (Figure 4). Such high correlation means that not only MDM fit general shapes and velocities of rotation curves, but also replicated all nuanced features of rotation curves such as local maxima and minima in velocities. Consistent with our theoretical observations, these results were achieved with mass distribution profiles typically showing rapid drop in density from a nucleus (median density of 5.9% at a distance 15% of $R_{max}$ from the nucleus), followed by a more gradual density decline in a galactic disk reaching median mass density of 0.1% at the optical edge of a galaxy.

The only instance where MDM automatic fit initially failed to achieve correlation over 0.995 was the case of galaxy NGC 5194. Specifically, the tail end of the rotation curve did not fit (Figure 4, dotted line), suggesting an unlikely high mass density at a galactic periphery. Further investigation[20,21] revealed that NGC 5194 has a nearby satellite galaxy NGC 5195 stealing matter from NGC 5194 periphery (Figure 2c). Once the adjustment for the satellite galaxy mass and distance was made, the rotation curve matched perfectly with correlation over 0.9999 (Figure 4). We believe this case illustrates MDM predictive and explanatory power, as it forces us to consider close galactic neighborhood that impacts the behavior of a particular galaxy.

## Figure 4 | Model fit to rotation curves of real galaxies

Name of each galaxy and estimated mass is shown on each chart. Blue circles represent observed rotation curve (data from http://www.ioa.s.u-tokyo.ac.jp/~sofue/); Red line denotes MDM generated expected rotation curve; and Green dashed line shows implied axisymmetric mass distribution. Left Y-axis shows speed in km/sec; Right Y-axis is mass density in % of maximum; and X-axis is distance from the nucleus in kpc. Includes 12 of total 47 galaxies; the remaining charts can be found in the extended figures.

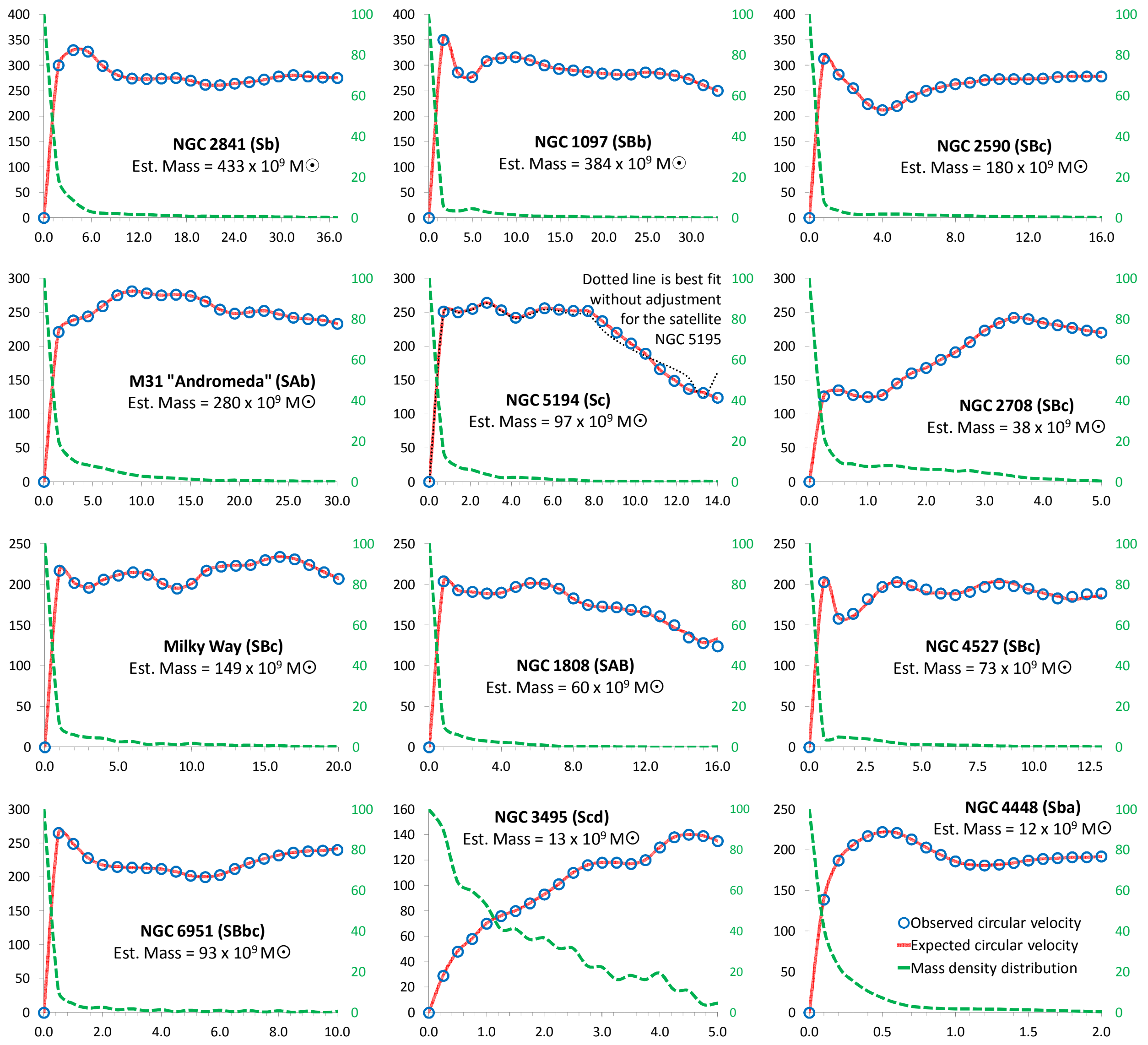

It is important to make a few observations about galactic masses. For the majority, even well researched galaxies, there is no consensus on galaxy mass between various sources. For example, the range of estimates for our close neighbor Andromeda galaxy M31 includes anything from $20*10^{10}$ M☉[16] to $1{,}230*10^{10}$ M☉[22]. Given that galaxy mass is not directly measured, but estimated from observed velocities on the basis of various galactic models, it is not surprising to see such broad range of estimates. As we explained above, our mass estimates and their respective distributions are computed automatically from our MDM model solving for the best fit to rotation curve. In the vast majority of cases, our mass estimates are below those commonly accepted. This leads to a conclusion that observed rotation curves can be readily explained without extra matter (dark or not) traditionally ascribed to galaxies.

TANGENTIAL GRAVITY FORCES IN SPIRAL GALAXIES

The additional advantage of MDM over earlier research methods is that MDM enables the analysis of tangential gravity forces and their impact on galaxy dynamics. In most cases tangential gravity force will be significantly weaker than centripetal force (Figure 5). This is due to the fact that sides of the galactic disk pull in opposite directions and largely cancel itself even with non-symmetric mass distribution. However, there are instances, when significant mass such as a stellar cluster is located close to the point of measurement, as is the case of galaxy M101 at 160 degrees axis (Figure 5b). In such situations, particularly on galactic periphery where centripetal force is weaker, the tangential force generated by nearby mass may exceed the centripetal force, significantly impacting the vector of the gravity force (Figure 5c) and consequently the shape of the rotation curve. This modeling of gravity forces at various axes allows us to make a testable prediction that rotation curves should differ in a spiral galaxy depending on the axis of measurement.

**Figure 5 | Tangential gravity forces in a spiral galaxy.**

**5a Model of Pinwheel galaxy M101: Axis at 0 degrees**

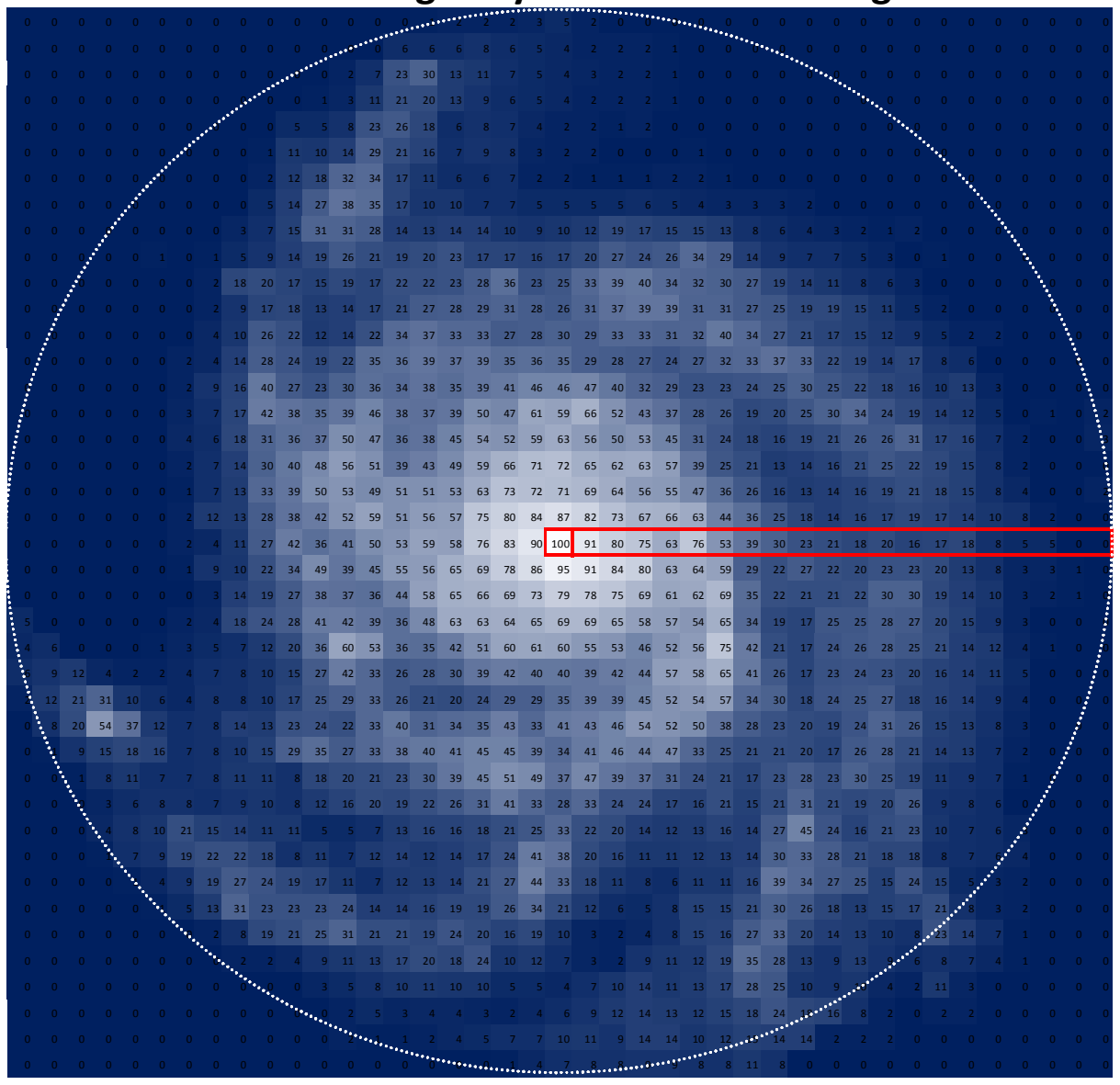

**5b Model of Pinwheel galaxy M101: Axis at 160 degrees**

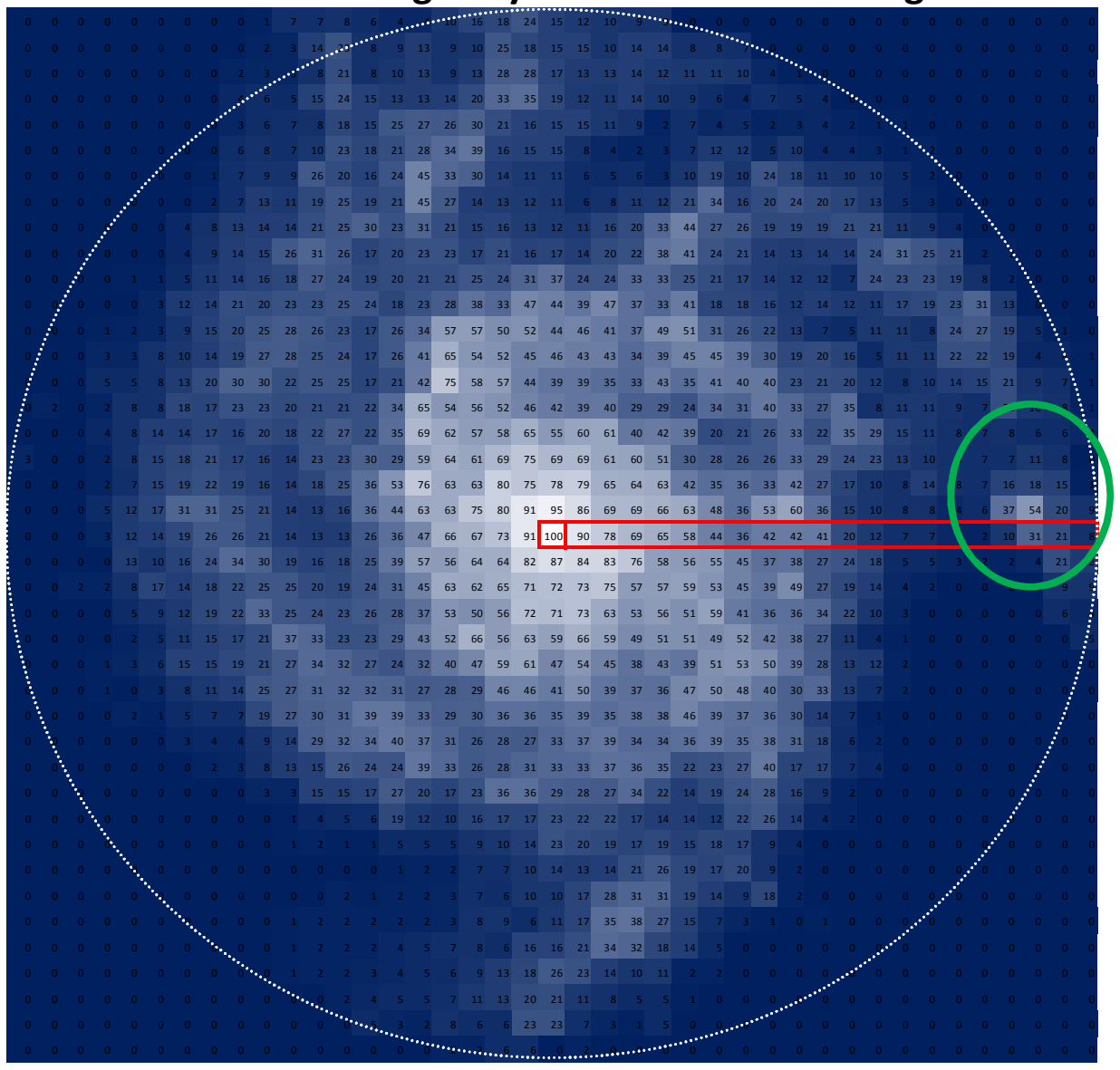

Legend: Axis of measument; Local mass generating strong tangential gravity force

**5c Centripetal and tangential gravity forces in M101 at 0 and 160 degrees axes**

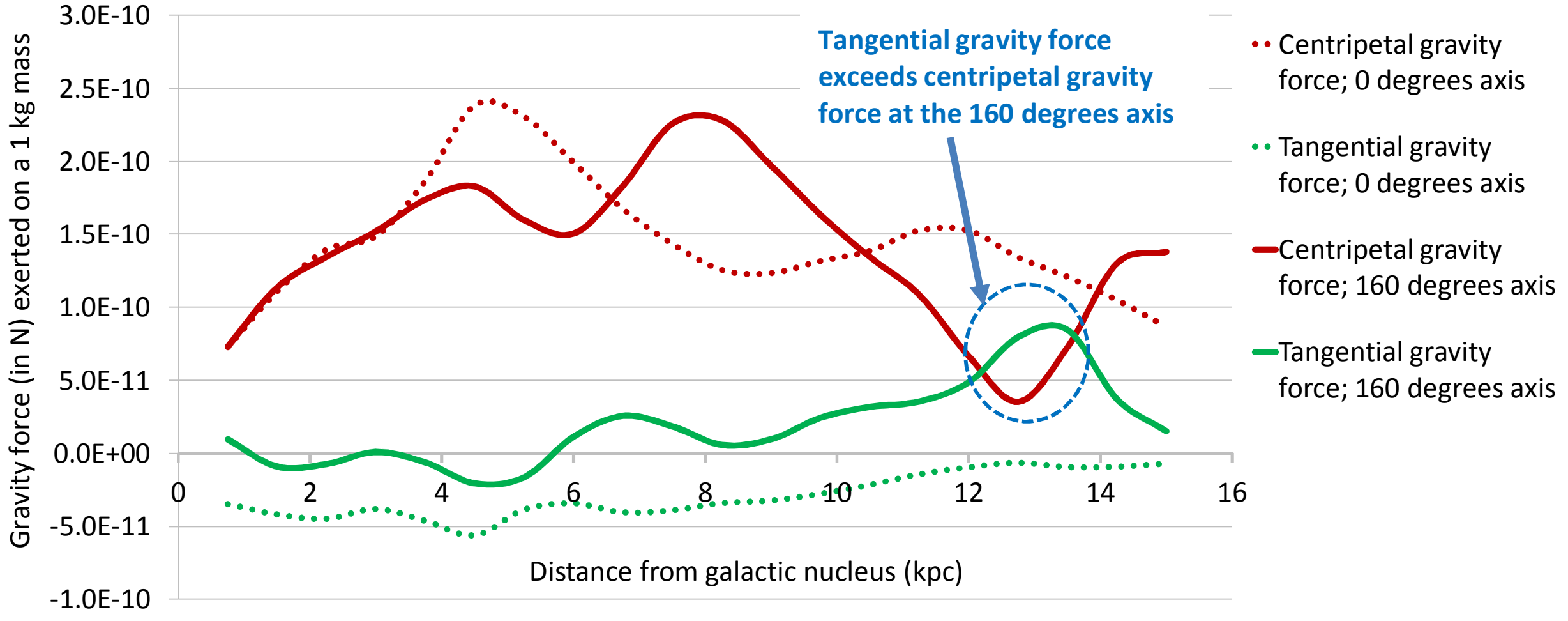

CONCLUSION

In this article we demonstrated that a broad spectrum of galaxy rotation curves can be explained without either dark matter or modified gravity, but by accounting for the distribution of baryonic matter and modeling its behavior strictly in accordance with the known law of gravity. Our Mass Density Model provides a theoretical framework for understanding of galaxy rotation and practical tool for modeling galaxy dynamics in relationship to its mass distribution. We also pointed two related mistakes of earlier research which led to the "invention" of dark matter and modified gravity theories—first, the expectation of Keplerian fall-off in circular velocities in galaxies similar to that in the Solar system, and, second, the assumption of the entire galaxy mass as a spherically symmetric gravitational field due to a point source for calculations inside the galaxy disk. Our results show that both are fundamentally incorrect.

ACKNOWLEGEMENTS

We want to sincerely thank Professor Yoshiaki Sofue of the University of Tokyo for making detailed rotation curve data on many galaxies publicly available.

EXTENDED FIGURES

See pages 10–12 below for extended figures.

## Extended Figure 1 | Model fit to rotation curves of real galaxies

Name of each galaxy and estimated mass is shown on each chart. Blue circles represent observed rotation curve (data from http://www.ioa.s.u-tokyo.ac.jp/~sofue/); Red line denotes MDM generated expected rotation curve; and Green dashed line shows implied axisymmetric mass distribution. Left Y-axis shows speed in km/sec; Right Y-axis is mass density in % of maximum; and X-axis is distance from the nucleus in kpc.

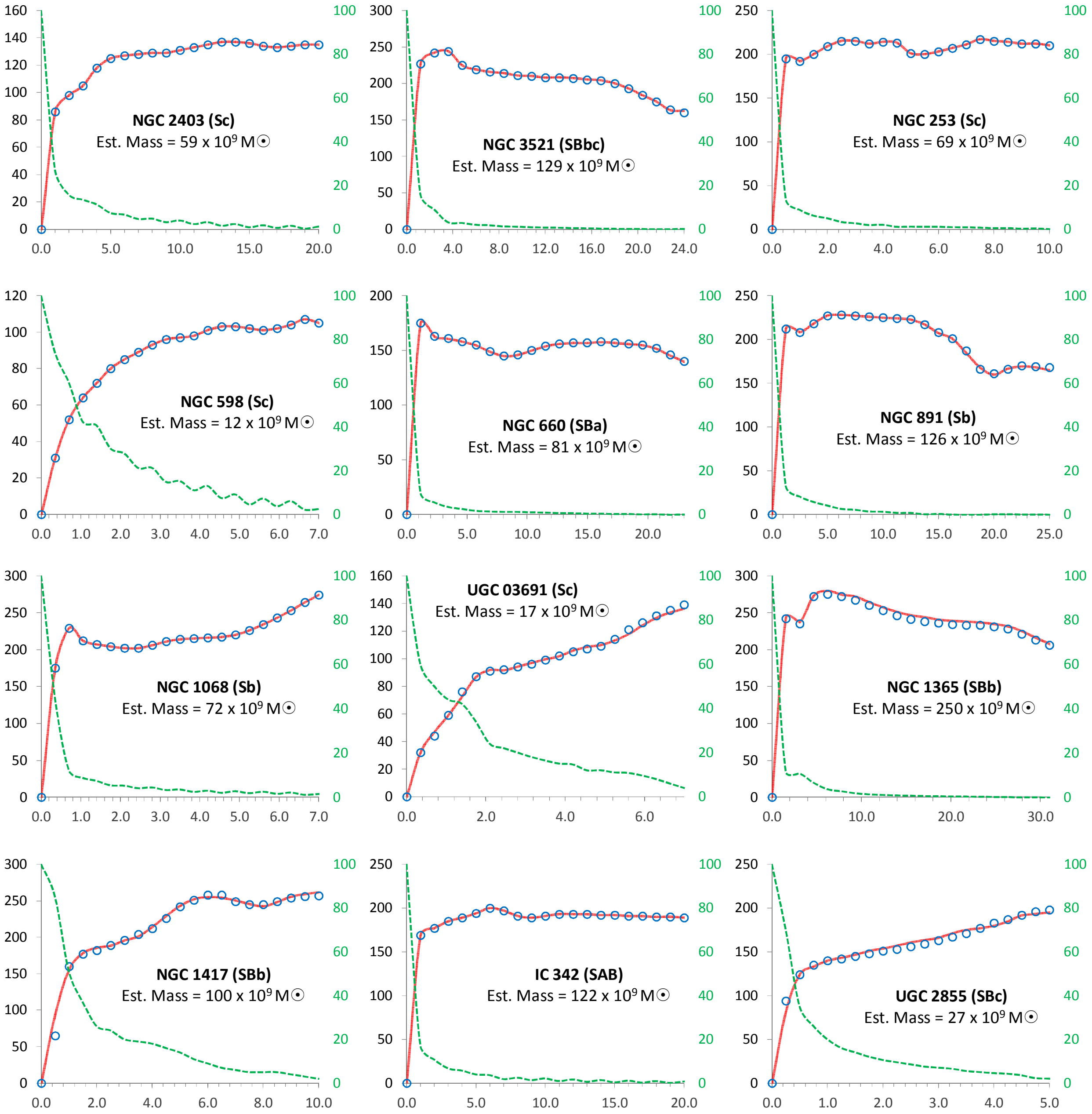

## Extended Figure 2 | Model fit to rotation curves of real galaxies

Name of each galaxy and estimated mass is shown on each chart. Blue circles represent observed rotation curve (data from http://www.ioa.s.u-tokyo.ac.jp/~sofue/); Red line denotes MDM generated expected rotation curve; and Green dashed line shows implied axisymmetric mass distribution. Left Y-axis shows speed in km/sec; Right Y-axis is mass density in % of maximum; and X-axis is distance from the nucleus in kpc.

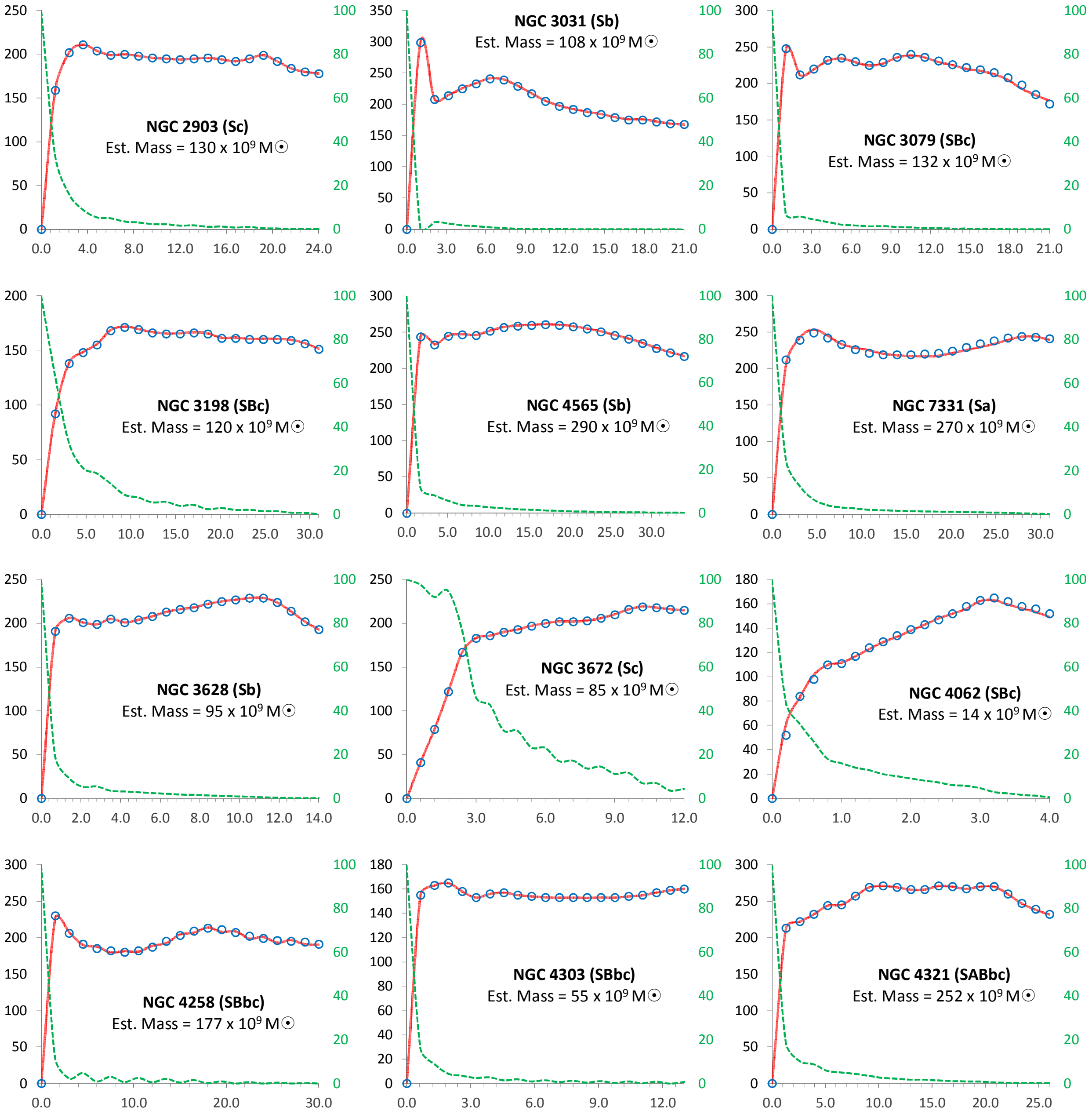

## Extended Figure 3 | Model fit to rotation curves of real galaxies

Name of each galaxy and estimated mass is shown on each chart. Blue circles represent observed rotation curve (data from http://www.ioa.s.u-tokyo.ac.jp/~sofue/); Red line denotes MDM generated expected rotation curve; and Green dashed line shows implied axisymmetric mass distribution. Left Y-axis shows speed in km/sec; Right Y-axis is mass density in % of maximum; and X-axis is distance from the nucleus in kpc.

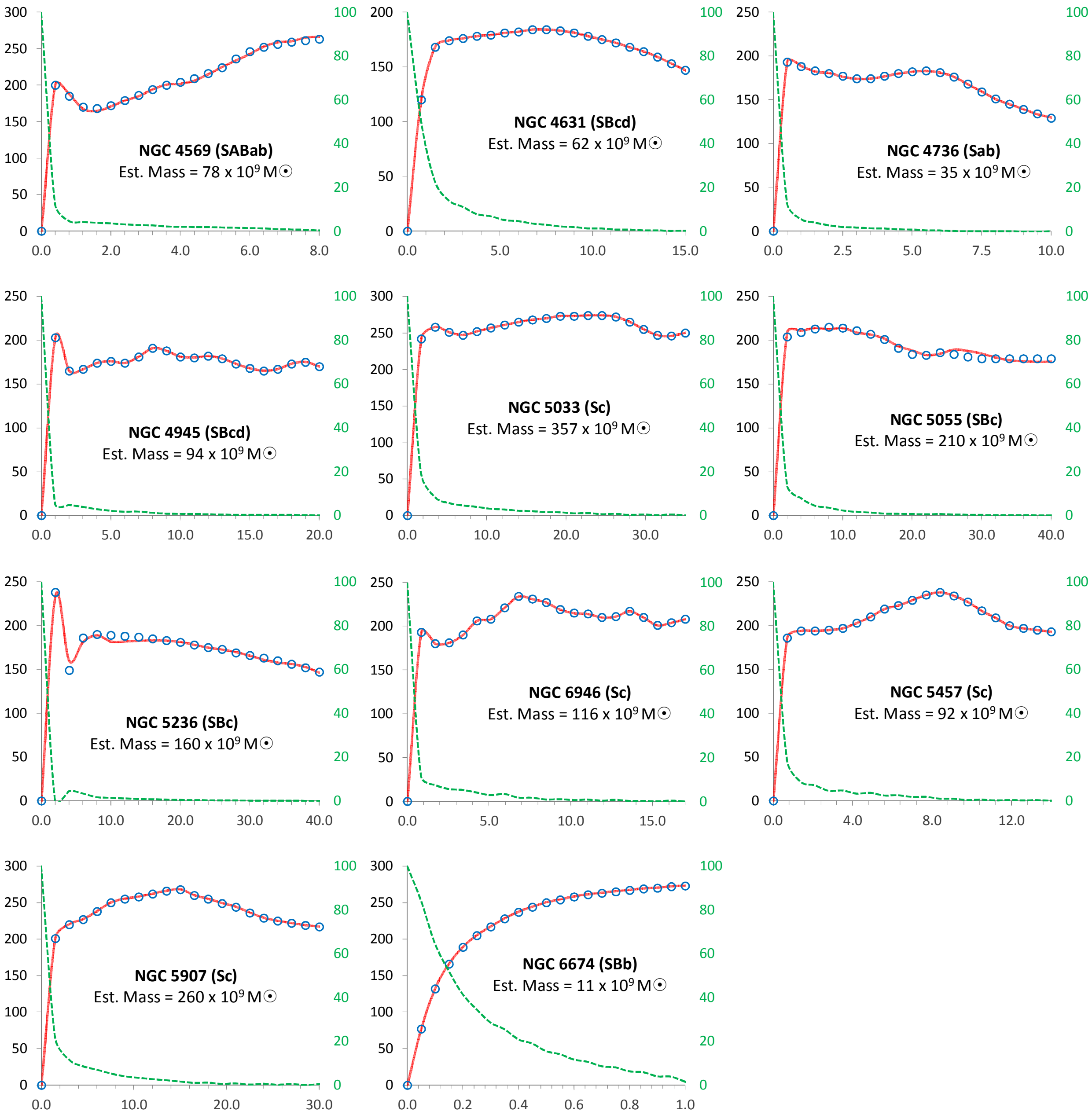